\documentclass[MSNbibl,nameyear,dvips]{arxstspdf}
\usepackage{flushend}
\usepackage{stfloats}

\volume{26}
\issue{2}
\pubyear{2011}
\firstpage{179}
\lastpage{184}
\referstodoi{10.1214/10-STS318}
\doi{10.1214/10-STS318A}

\makeatletter
\def\bsuffix #1{#1}
\makeatother

\begin{document}
\begin{frontmatter}

\title{Discussion of ``Calibrated Bayes, for Statistics in General, and Missing Data
in Particular'' by R.~J.~A.~Little}
\runtitle{Discussion}
\pdftitle{Discussion of Calibrated Bayes, for Statistics in General, and Missing Data
in Particular by R.~J.~A.~Little}

\begin{aug}
\author[a]{\fnms{Nathaniel} \snm{Schenker}\corref{}\ead[label=e1]{nschenker@cdc.gov}}

\runauthor{N. Schenker}

\affiliation{Centers for Disease Control and Prevention}

\address[a]{Nathaniel Schenker is Associate Director for Research and
Methodology,
National Center for Health Statistics,
Centers for Disease Control and Prevention,
3311 Toledo Road,
Hyattsville, Maryland 20782, USA
\printead{e1}}

\end{aug}



\end{frontmatter}

It is a pleasure and an honor for me to comment on this article by Rod
Little, who has contributed greatly to statistics in general and to
Bayesian statistics and handling missing data in particular. Little
provides a nice discussion of the calibrated Bayes approach, methods for
missing-data problems and recent developments (SRMI and PSPP) that
increase flexibility in dealing with missing data.

\section{Don't Forget the Pragmatists}\label{1}

Little begins his Section 2 by stating that the statistics world is
still largely divided into frequentists and Bayesians. Indeed, during
the University of Maryland workshop (``Bayesian Methods that
Frequentists Should Know'') at which Little presented a talk on the
topic of his article, many of the speakers declared themselves to be
either frequentists or Bayesians. As formal discussant of Little's talk,
however, I declared myself to be a ``pragmatist,'' which Little (\citeyear{Lit06})
defined as one who does not have an overarching philosophy and picks and
chooses what seems to work for the problem at hand. If I were forced to
choose a philosophy, I would probably go with the Bayesian one. But I am
happy to use either approach, depending on the context, and many of my
statistical colleagues seem willing to use either approach as well.
Moreover, although subject-matter specialists with whom I work seem to
be primarily familiar with point estimates, standard errors and
confidence intervals, they seem to have no problems using Bayesian
analogues (e.g., posterior means, standard deviations and credibility
intervals) in the same way, when presented with them.

Little (\citeyear{Lit06}) argued that, to enhance the credibility of our profession
and avoid confusion and ambiguity, it would be preferable not to have
the ``split personality'' that is inherent in the pragmatic approach. He
has made a strong case in that article and here for calibrated Bayes as
a unified inferential approach that combines strengths of the Bayesian
and frequentist approaches. His arguments are compelling, but given the
abundance of good and easily accessible frequentist methods that exist
and are widely used, I imagine that it would be difficult for our
profession to rid itself of this split personality. Moreover, I think
the key issue in most applications is the development of realistic
models for the data. Thus, I second Little's emphasis on flexible models
and methods, such as the SRMI and PSPP methods, and his concluding call
for further work on model diagnostics, especially in the area of missing
data.

\begin{table*}
\caption{Simplified, nonexhaustive 2-way table depicting the
frequentist/Bayesian dichotomy within survey~sampling~and~in~many~areas~outside~of~survey~sampling}
\label{tab1}
\begin{tabular*}{\textwidth}{@{\extracolsep{\fill}}lll@{}}
\hline
\textbf{Mode of inference} & \multicolumn{1}{c}{\textbf{Within survey sampling}} & \multicolumn{1}{@{\hspace*{-1pt}}c@{}}{\textbf{In many areas outside of survey sampling}}\\
\hline
Frequentist & $\bullet$ Estimate $Q(Y)$ by $\hat{Q}(Y_{\mathrm{inc}},I)$, where $Y =$ population values, & $\bullet$ Formulate $p(y|\theta)$, where $y =$ data and \\
&$Y_{\mathrm{inc}} =$ sampled values, and $I =$ indicators of inclusion &$\theta =$ parameters.\\
&in the sample.&\\
 & $\bullet$ Base inferences for $Q(Y)$ on $p(\hat{Q}(Y_{\mathrm{inc}},I)|Y)$ induced& $\bullet$ Base inferences for $\theta$ on $p(\hat{\theta} (y)\vert\theta)$,\\
& by the distribution of the indicators $I$ in repeated& where $\hat{\theta}( y)$ estimates $\theta$.\\
& sampling, $p(I|Y)$. &\\[5pt]
Bayesian & $\bullet$ Formulate $p(Y|\theta)$ and $p(\theta)$, in addition to $p(I|Y)$. & $\bullet$ Formulate $p(y|\theta)$ and $p(\theta)$.\\
 & $\bullet$ Base inferences for $Q(Y)$ on $p(Q(Y)|Y_{\mathrm{inc}},I)$. & $\bullet$ Base inferences for $\theta$ on $p(\theta |y)$.\\
\hline
\end{tabular*}
\end{table*}

\section{The Frequentist/Bayesian Schism Is Perhaps Magnified in
Survey Sampling}\label{2}

In the survey sampling world in which I primarily work as a government
statistician, the definition of being a frequentist versus being a
Bayesian is not necessarily clear, because inferences are often desired
about finite-population quantities rather than about model parameters.
Such inferences are often made using a design-based paradigm (e.g.,
Cochran, \citeyear{Coc77}), that is, based on the distribution of estimators in
repeated sampling from the finite population under a given design. Thus,
one possible definition of frequentist inference in survey sampling is
that it treats the finite-population values, $Y$, as fixed parameters,
and bases inferences about a function of those parameters, say, $Q(Y)$,
on a function of the sampled values and its distribution in repeated
sampling. The corresponding definition of Bayesian inference (e.g.,
Rubin, \citeyear{Rub87}, Chapter 2) is that it places a prior distribution on $Y$,
say, $p(Y|\theta)$, where $\theta$ represents hyperparameters with a
hyperprior $p(\theta)$, and bases inferences on the posterior predictive
distribution of $Q(Y)$ given the sampled values.

The two-by-two table (Table \ref{tab1}) gives a simplified, nonexhaustive depiction
of the frequentist/Bayesian dichotomy within survey sampling on the one
hand and in many areas outside of survey sampling on the other. As
Table \ref{tab1} shows, both within and outside of survey sampling, there are
differences between the frequentist and Bayesian approaches concerning
which quantities are treated as random, as well as whether prior
distributions are specified. However, within survey sampling, there is
an additional distinction, which is perhaps the most important in
practice. The reference distribution for inferences under the
frequentist, or design-based approach, is not induced by a model for the
finite-population values, $Y$, whereas the Bayesian posterior predictive
distribution does involve such a model.

Much has been written on design-based versus model-based inference in
sample surveys, but I would particularly like to cite Hansen, Madow and
Tepping (\citeyear{HANMADTEP83}) and Little (\citeyear{Lit04}). Hansen, Madow and
Tepping (\citeyear{HANMADTEP83}) concluded
basically that for descriptive inference from reasonably large,
well-designed sample surveys, design-based inference is to be preferred,
because it avoids errors due to model misspecification that are possible
with model-based inference, and because it loses little efficiency
relative to model-based inference. They acknowledged, however, that
model-based methods for sample surveys can be useful and important in
the contexts of sample design, inference for small samples, inference in
the presence of nonsampling errors, and situations in which inferences
under a model are of intrinsic\vadjust{\eject} interest. One issue regarding the
conclusions of Hansen, Madow and
Tepping (\citeyear{HANMADTEP83}) is that it is not always clear how
large a sample is large enough. Moreover, lately there has been
increasing interest in ``pushing the data as far as possible,'' for
example, by using a national survey to obtain estimates for a small
subpopulation.

$\!\!$Little (\citeyear{Lit04}) concluded that the Bayesian paradigm is flexible enough to
provide practical and useful inferences in the context of survey
sampling. He pointed out that the models used in Bayesian inference for
surveys need to properly reflect features of the sample design, such as
weighting, stratification and clustering, or else inferences are likely
to be distorted. Similar points were made in the discussions of Hansen, Madow and
Tepping (\citeyear{HANMADTEP83}), in particular, those by Rubin, who clarified the role of
the probabilities of selection in Bayesian modeling for sample surveys,
and Little, who advocated the use of model-based estimators that are
design-consistent. Hansen, Madow and
Tepping (\citeyear{HANMADTEP83}) agreed with those points in
their rejoinder. However, as I will discuss in Section \ref{43} in the
context of applications to be presented in the next section, reflecting
sample design features can be complicated in some problems for which
model-based inference can be particularly useful. Thus, I believe that
further development of methods for reflecting design features will be an
important area of research.

\section{A Major Reason Why This Pragmatist Likes Bayesian Methods}\label{3}

From a pragmatic point of view, one of the major attractions of Bayesian
methods is their ability to handle problems with complex data structures
such as missing data in a relatively straightforward manner. As Little
points out, this has been true\vadjust{\eject} especially since the development of
Markov chain Monte Carlo methods and multiple imputation. To complement
Little's discussion and to illustrate some of his points, I will now
describe a few applied projects for which Bayesian techniques were very
helpful.

\subsection{Survival Analysis with Intermittently Observed Covariates}\label{31}

Faucett, Schenker and Elashoff (\citeyear{autokey3}) analyzed the relationship between
post-operative smoking and survival using data from a clinical trial on
survival of patients after surgery for lung cancer. At follow-up visits,
the patients had been asked about their current smoking status. Faucett, Schenker and Elashoff (\citeyear{autokey3}) discretized time using narrow time intervals, and they
specified a Markov chain model for current smoking status together with
a time-dependent proportional hazards model with a piecewise constant
baseline hazard for survival given smoking behavior and covariates.
Gibbs sampling was used to approximate the joint posterior distribution
of the model parameters under diffuse prior distributions.

The use of Gibbs sampling facilitated analyses under two different
survival models, one with current smoking as the time-dependent
covariate and another with cumulative smoking as the time-depen\-dent
covariate. It was found that the coefficient for cumulative smoking (in
the latter model) had much more posterior probability mass to one side
of zero than did the coefficient for current smoking (in the former
model). Thus, the evidence was stronger for a~detrimental effect of
cumulative smoking than for a~detrimental effect of current smoking. The
application of Faucett, Schenker and Elashoff (\citeyear{autokey3}) is an example of joint modeling of
longitudinal and survival data, which has been a popular area of
research in the past decade.

\subsection{Incorporating Auxiliary Variables into Survival Analysis
via Multiple Imputation}\label{32}

In a different type of application that jointly modeled longitudinal and
survival data, Faucett, Schen\-ker and Taylor (\citeyear{FauSchTay02}) developed an
approach, based on multiple imputation, to using auxiliary variables to
recover information from censored observations in survival analysis.
Applications of this type are mentioned by Little in his Section 5,
point (a) and Section 7. Faucett, Schenker and Taylor (\citeyear{FauSchTay02}) analyzed data from an AIDS
clinical trial comparing zidovudine and placebo, in which the outcome of
interest was the time to development of AIDS, and in which CD4 count was
a time-dependent auxiliary variable. Because AIDS can take a long time
to develop, most of the observations were censored.
Faucett, Schenker and Taylor (\citeyear{FauSchTay02})
specified a~hierarchical change-point model for CD4 counts and a~%
time-dependent proportional hazards model for the time to AIDS given CD4
and covariates. Markov chain Monte Carlo methods were then used to
multiply impute event times for the censored cases.

The use of multiple imputation facilitated drawing inferences about
quantities whose posterior distributions could not be approximated
directly using the output of the Markov chain Monte Carlo simulations.
For example, Kaplan--Meier estimates of survival under treatment and
placebo were compared, and the coefficient of treatment in a Cox
regression analysis was \mbox{examined} as well. Comparisons with analyses of
the censored data without imputation, and accompanying simulation
results, suggested that incorporating the auxiliary variables via
multiple imputation can lead to improved efficiency as well as partial
corrections for dependent censoring. This application illustrated use of
a nonBayesian complete-data analysis with multiple imputation; see
Little's Section 5, point (c).

\subsection{Multiple Imputation for Missing Data in Surveys}\label{33}

As Little discusses in Section 5, points (a) and (b), multiple imputation
has particular benefits in the context of public-use data. SRMI was used
recently in two major applications of multiple imputation to public-use
data from the National Center for Health Statistics. One involved
missing income data in the National Health Interview Survey (NHIS)
(Schenker et al., \citeyear{Schetal06}), and the other involved missing body-scan data
from dual-energy X-ray absorptiometry (DXA) in the National Health and
Nutrition Examination Survey (NHANES) (Schenker et al., \citeyear{SCHBOR}). DXA
scans are used to measure body composition such as soft tissue
composition and bone mineral~con\-tent. Public-use data with multiple
imputations from both applications have been released online
(\href{http://www.cdc.gov/nchs/nhis/2009imputedincome.htm}{http://}
\href{http://www.cdc.gov/nchs/nhis/2009imputedincome.htm}{www.cdc.gov/nchs/nhis/2009imputedincome.htm};\break
\href{http://www.cdc.gov/nchs/nhanes/dxx/dxa.htm}{http://www.cdc.gov/nchs/nhanes/dxx/dxa.htm}).

Both applications involved nontrivial amounts of missing data---roughly
30\% for the NHIS income data and 20\% for the NHANES DXA data---with
missingness related to characteristics of the persons surveyed, so that
analysis of only the complete cases would likely result in biases as
well as inefficiencies. The use of SRMI facilitated inclusion of large
numbers of predictors of different types (e.g., categorical, continuous,
count) in each application, with some of the predictors having missing
data themselves, although usually at much lower levels than the main
variables of interest. As discussed in Meng (\citeyear{MEN94}), Rubin (\citeyear{RUB96}), Little
and Raghunathan (\citeyear{LITRAG97}) and Reiter, Raghunathan and
  Kinney (\citeyear{REIRAGKIN06}), the reasons for
including several predictors in imputation models, besides of course to
help predict the missing values, are to help ``explain'' the
missingness, that is, to make the assumption of missingness at random
more tenable, and to promote compatibility between the imputation models
and the analyses that would ultimately be carried out by secondary users
of the data. The issue of possible incompatibility is discussed further
in Section \ref{44}.

Each application had other interesting features, some of which were
handled especially well by SRMI. For example, in the NHIS project, for
the \mbox{majority} of the missing family income values, respondents~had
provided coarse income categories, so that bounds were available for the
missing values. Also, there~we\-re sometimes structural dependencies
between varia\-bles that needed to be imputed. For example, a~person could
not have earnings unless he/she was emp\-loyed, and occasionally,
employment status was mis\-sing along with earnings.

In the NHANES project, the missing data were highly multivariate. There
were 32 DXA variables, some of which were highly interrelated; and
sometimes the DXA data were only partially missing. Perhaps the most
interesting feature of the project, however, was that missingness of DXA
data often occurred for people with high levels of truncal adiposity,
because the adiposity interfered with the ability to obtain valid
measurements. Thus, the levels of missingness tended to be high at the
largest values of other variables measured in the NHANES, such as BMI
and waist circumference. This necessitated some extrapolation beyond the
range of the observed DXA values.

\subsection{Combining Information from Two Surveys to Enhance
Small-Area Estimation}\label{34}

A multi-organization project led by the National Cancer Institute used
Bayesian methods to compute small-area estimates of the prevalence of
cancer risk factors and cancer screening by combining information from
two surveys for the years 1997--2003 (Raghunathan et al., \citeyear{Ragetal07}; Davis et
al., \citeyear{Davetal}). The surveys were the Behavioral Risk Factor Surveillance
System (BRFSS), a large, state-based survey conducted by telephone, and
the NHIS, a smaller, face-to-face survey. The BRFSS included most of the
counties in the United States in its sample and thus provided some
direct information about them. However, it obtained data only from
households equipped with telephones, and its nonresponse rates tended to
be relatively high, as is often the case with telephone surveys. The
NHIS surveyed both telephone and nontelephone households, asked a
question to identify the telephone status of the household, and
generally had lower nonresponse rates than the BRFSS. However, its
sample only included about 25 percent of the counties.

A Bayesian, trivariate extension of the Fay--Herriot (\citeyear{FayHer79}) model was
formulated. Markov chain Monte Carlo methods were used, together with
county-level telephone coverage rates from the 2000 census, to
approximate the posterior distributions of the small-area rates.
Estimates from the project have been released publicly
(\url{http://sae.cancer.gov/}).

\section{Some Areas for Further Research}\label{4}

\subsection{Flexible Models and Methods}\label{41}

In Section \ref{1} I mentioned the need for more flexible models and methods.
SRMI and PSPP are two examples of techniques that have increased
flexibility (see, e.g., Section \ref{33} for examples in which SRMI was
used), and the development of more such techniques would be welcome. For
example, perhaps a~flexible univariate prediction model such as PSPP
could be used for each univariate regression in SRMI to develop a robust
procedure for multivariate imputation.

\subsection{Diagnostics for Models}\label{42}

In Section \ref{1} I also seconded Little's call for work in the area of model
checking, especially for missing-data problems. With missing data,
checking prediction models for the missing values is especially
difficult, for the obvious reason that the missing values are
unavailable for use in model checking. Diagnostics for imputations of
the general types mentioned in Abayomi, Gelman and Levy (\citeyear{AbaGelLev08}) were used in the
NHANES multiple-imputation project discussed in Section \ref{33}.

Little (Section 2) mentions methods such as posterior predictive checks
as being frequentist in spirit. It would be helpful to investigate more
fully the link between use of such techniques and achieving
well-calibrated analyses, such as Bayesian credibility intervals with
good frequentist coverage properties. Also useful for survey
practitioners would be more research on evaluating models from a
design-based point of view, especially in the context of complex sample
designs.

\subsection{Incorporating Complex Sample Design Features into Models}\label{43}

As I mentioned in Section \ref{2}, incorporating complex sample design
features into models for survey data can be complicated. In the context
of multiple imputation, inclusion of survey weights and indicator
variables for strata and primary sampling units (PSUs) has been
advocated (Rubin, \citeyear{RUB96}; Reiter, Raghunathan and Kinney, \citeyear{REIRAGKIN06}). Such
techniques were used in the NHIS and NHANES multiple-impu\-tation projects
described in Section \ref{33} above, although in the NHANES project, there
was some concern about parsimony, so a smaller number of variables
related to PSU selection were substituted for the full set of indicator
variables. Further work on methods for increasing parsimony, such as via
use of random effects, would be helpful.

In addition, incorporating complex sample design features can be
difficult in problems that involve~com\-bining information across surveys,
because the \mbox{design} features of the two surveys might not be comparable.
This was one reason for using an area-level (Fay--Herriot) rather than
person-level model in the small-area estimation project discussed in
Section~\ref{34}; see Schenker and Raghunathan (\citeyear{SCHRAG08}). Schenker, Raghunathan
and Bondarenko (\citeyear{SCHRAGBON10}) also discussed such issues in the context of using
multiple-imputation to combine information from two surveys.

\subsection{Impacts of Secondary Analysts Using Variables not Included
in the Imputation Model}\label{44}

Little notes (Section 5) that an attractive feature of multiple
imputation is that the imputation model can include variables not
included in the final analysis. I agree with this, and, furthermore, I
have found multiple imputation to be a very general and flexible method
for allowing secondary analysts of public-use data to assess the
uncertainty due to imputation.

A concern of mine, which applies to single imputation as well as
multiple imputation, is biases that can occur in point estimates of
interest when a secondary analyst uses the imputed data together with
variables that were not included in the imputation model. As mentioned
in Section \ref{33}, the NHIS and NHANES projects used large numbers of
predictors in order to avoid such incompatibilities, and the predictors
were listed for secondary analysts in the technical documentation for
the projects. However, it is likely in general that some secondary
analysts of public-use data will attempt analyses that ``go beyond'' the
imputation model. The biases in point estimates for such analyses will
depend in a sense on how well the variables included in the imputation
model account for the relations being studied in the secondary analysis.
Further research on the possible extent of such biases, and guidelines
and diagnostics for secondary analysts, would be useful areas for
research.

\subsection{Real-Life Examples of the Utility of the Calibrated Bayes
Approach}\label{45}

As I mentioned in Section \ref{1}, I imagine that it would be difficult to
move our field completely away from having a ``split personality'' and
toward following Little's (\citeyear{Lit06}) ``Bayes/Frequentist Roadmap.''
Excellent papers such as Little's current one will provide nudges in
that direction. Also helpful will be more real-life examples of how the
calibrated Bayes approach can help to achieve substantial gains in
solving problems that could not be achieved otherwise.

\section*{Acknowledgments}
The author thanks John Eltinge, Jennifer
Madans, Van Parsons and the guest editors for helpful comments on
earlier drafts of this discussion. The findings and opinions expressed herein are those of the author and do not necessarily
reflect the views of the National Center for Health Statistics, Centers for Disease Control
and Prevention.


\end{document}